# Component-based Situational Methods
*A framework for understanding SME*

Yves-Roger Nehan – Rébecca Deneckere
CRI, University Paris 1 - Panthéon Sorbonne
90, rue de Tolbiac, 75013 Paris, France,
Habas.Nehan@malix.univ-paris1.fr,rebecca.deneckere@univ-paris1.fr
WWW home page: http://crinfo.univ-paris1.fr/

**Abstract.** The work presented in this paper is related to the area of Situational Method Engineering (SME) which focuses on project-specific method construction. We propose a faceted framework to understand and classify issues in system development SME. The framework identifies four different but complementary viewpoints. Each view allows us to capture a particular aspect of situational methods. Inter-relationships between these views show how they influence each other. In order to study, understand and classify a particular view of SME in its diversity, we associate a set of facets with each view. As a facet allows an in-depth description of one specific aspect of SME, the views show the variety and diversity of these aspects.

## 1. Introduction

Method Engineering aims to bring effective solutions to the construction, improvement and modification of the methods used to develop information and software systems. Several authors tried to design methods that would be as effective and as adapted as possible to the development needs of information systems [1,2]. This goal was not always reached, especially because the methods were not always well adapted to projects specificities. The situational methods were designed to correct this weakness. The situational approach finds its justification in the practical field analysis which shows that a method is never followed literally [3, 4]. The discipline of Situational Method Engineering (SME) promotes the idea of retrieving, adapting and tailoring components, rather than complete methodologies, to specific situations [5]. In order to succeed in creating good methodologies that best suit given situations, components (building blocks of methodologies) representation and cataloguing are very important activities. In particular, the components have to be represented in a uniform way that includes all the necessary information that may influence their retrieval and assembling. This paper is an attempt to explore some of the issues underlying component-based approaches to Situational Method Engineering (SME) and to propose a framework for their classification. This framework is 4-dimensional as it advocates that a SME approach can be defined by four views, each capturing a particular aspect of SME. Each view has multiple facets and the associated metric. The idea of a four views framework and its facets has been used in several domains such as: requirements engineering for understanding and classifying scenario based approaches [6], system engineering [7], etc.

When used in the SME domain, a facet provides a means of classification. For instance, the formalism facet of the system view (see section below) helps in classifying SMEs according to the underlying paradigm used: informal, semi-formal and formal. Each facet has values which are defined in a domain. A domain may be a predefined type, an enumerated type, or a structured type.

We use the four views framework as a baseline and attach an aspect of SME to each of the views and a set of facets to each view. As a result, it is possible to identify and investigate four major viewpoints of SME: what is the objective of SME, , how are represented the method components, how can the methods be developed and used and finally what does SME achieve.

This paper is organised as follows: Section 2 describes our four views framework. Section 3, 4, 5, 6 explain each view and list a set of their facets for comparing and evaluating the component representation approach. Section 7 presents and illustrates eight of the most recent situational methods, then further analyses each SME approach according to these four different views of our framework. A conclusion is done in Section 8.

## 2. The Four-views Framework

The four views framework, originally proposed in [7], has proved its efficiency in enhancing the understanding of various engineering disciplines such as information systems engineering [7], requirements engineering [8], IS development process engineering [9] and method engineering [10].

In the original SE framework [7], the views where described as follows.

- The subject view contains knowledge of the domain about which the proposed IS has to provide information. It contains real-world objects which become the subject matter for system modeling.
- The system view includes specifications of what the system does, at different levels of detail. It holds the modeled entities, events, processes, etc. of the subject world as well as the mapping onto design specifications and implementations.
- The usage view describes the organizational environment of the information system, i.e. the activity of agents and how the system is used to achieve work, including the stakeholders who are system owners and users.
- The development view focuses on the entities and activities which arise as part of the engineering process itself.

Our point of view is that this framework concept can be used to help in understanding the field of SME disciplines which consists of applying engineering approaches, techniques, and tools to the construction and representation of components. The purpose of this work is then to present a state of the art in Situational Method Engineering. The four views composing the 4-dimensional framework proposed in this work try to answer the following questions about component-based situational methods:
- what is a component-based situational method ?
- how is represented a situational method component ?
- how can situational methods be developed and used ?
- what is the rationale of component-based situational method engineering ?

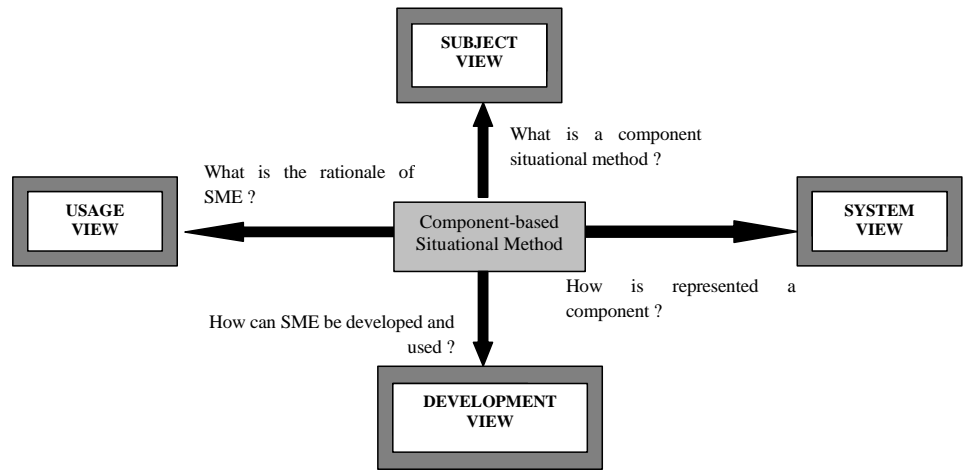

**Fig. 1**: The four views of SME

For our purpose, we define the SME 4-dimensional framework as follows.
- The subject view as the dimension which deals with the situational method definition, its nature.
- The representation of method component is described in the system view.
- In the usage view, we will investigate the reasons, the rationale for SME and relate users needs to the situational methods that can best meet them.
- The development view deals with the process of constructing component-based situational methods. This process is a meta-process in that it supports the construction of components which will in turn support the development of methods. The way this process might be supported by a tool environment is also relevant in this view.

This allows us to discuss in a focused manner the different concerns of SME: the definitions of components, their representations, the way of developing these representations, and the rationale for using these representations. This is done in the subject, system, development, and usage views respectively. Each view is described by a set of facets that allow a more detailed study of the situational methods. However, this set is not exhaustive and can be completed by other studies.

## 3. The Subject view

This view of SME deals with the notion of method nature.

In [11], I. Mirbel defines different objectives that are targeted by the approaches. Following this typology, we define a facet representing the nature of the SME methods.
- A first family of approaches aims at *documenting methods* through well-defined components [12, 13]. This kind of method does not state precisely how to retrieve and reuse a component but offers a good effort of specification with regards to the elements a method is made of.
- The second category focuses on the *retrieving* of components to reuse them and *evaluating* their similarity [3].
- The third category focuses on method fragmentation with the definition of *guidelines for reusing* the different components in daily developer tasks by project team members [4,14].

The nature of SMEs can thus be classified as follows:
*Nature*: SET (ENUM {Documenting, Retrieving and evaluating, Reuse guiding})

## 4. The Usage view

The usage view concerns the objectives we try to achieve with SME methods as well as the means necessary to their implementation. The SME approaches use high flexibility and thus modify methods to adapt them to a given situation by taking account of its specificities. This leads us to see the usage view as imposing three strict requirements : how the methods must be constructed the nature of method components and how these components must be developed. It is in the usage view that the method objectives must be stated. These aspects depend on the components management policy. This policy is to build methods starting from components whose names and contents vary according to the design. The use of components allows capturing knowledge which changes with time. The use of a library allows capitalizing the experiments of prior projects. A particular policy may be formalised with the two following facets: C*onstruction technique* and *knowledge representation*.

### 4.1 Construction technique facet

This facet represents the various ways of building a method which are instantiation, assembly, extension and reduction.
- *The instantiation approaches* use an identification of the common and generic method characteristics and represent them by a system of concepts called meta-model. These approaches allow the creation of a whole set of methods sharing the same properties [3, 15, 16].
- *The assembly approaches* concentrate on the grouping of method components belonging to complementary methods [3,17]. They assemble separate selected method components with regard to the studied specific project to form a unique method. To be successful, it is necessary to have a modular process model.
- *The extension approaches* allow the transformation of a basic method into a new method adapted to the project's needs [18, 19] with addition of new functionalities in a base method.
- *The reduction approaches* allow removal of basic method operators in order to transform it to match the engineer's needs [20, 21].

A method can be classified according to its defined type of construction:
*Construction technique*: SET (ENUM {instantiation, assembly, extension, reduction})

### 4.2 Knowledge representation facet

The question of the component retrieval is an important issue of the SME field. Three possibilities have emerged in the literature : (1) the project is globally characterized with use of contingency factors, (2) the components are described with use of descriptors and (3) patterns are used to instantiate the right componant following the project needs. We can then define three SME categories following the knowledge representation.

- *An SME fragment based method* consists in encouraging a global analysis of the projects while basing itself on contingency criteria. The projects and the situations are characterized by means of factors associated with the methods. [22] uses a contingency model based on 17 contingency factors which take value between Low and High as 'Importance of the Project', 'Knowledge and Experience', 'Stability' and so on. According to the authors, the characterization of the project allows them to select the method components appropriate to the project. Construction is supported by component assembly rules and constraints having to be satisfied by the created method.

- *An SME chunk based method* aim at associating these reusable components to their description in order to facilitate component research and extraction according to the user's needs. [3] uses the concept of descriptor [23] like a means to describe method components. The descriptors are organized in a contextual way: each one of them defines the situation in which the component can be employed and describe its usage intention.

- *An SME pattern based method*. A pattern describes a recurring problem with his associated solution [18]. It provides a solution which becomes reusable for any situation concerned with this problem. By developing patterns, the users condense part of their knowledge on the field of the problem and allow its availability for the other users.

The knowledge representation can thus be classified as follows:
*Knowledge representation* : ENUM {fragment, chunk, pattern}

### 5. The System view

This specific view is focused on the component representation by defining what is represented, at what level of abstraction, how is it represented and what properties should have the representation. These aspects are captured by the following three facets : Dimension, Abstraction and Formalism.

### 5.1 Dimension facet

A component is not always viewed with the same dimension. The situational methods use various techniques to represent knowledge: fragments, chunks and patterns. Although terminology between research groups differs, typically a chunk [3, 6] will encapsulate both a process and a product part whereas a fragment [24] can be either a product or a process fragment:
- *product* fragment relate to the structural and static aspects of methodologies (e.g.; deliverables, documents, models, diagrams, and concepts), whereas
- *process* fragment capture the behavioral and procedural aspects of methodologies (stage, tasks, activities, and techniques to be carried out) [25].

Dimension can thus be classified as follows:
*Dimension*: SET (ENUM {product-oriented, process-oriented})

### 5.2 Abstraction facet

In [24], this notion in SME is related to the abstraction level of a component that can be :
- *conceptual*, as in [20] where components are expressed with descriptions and specifications of methodology parts, or
- *technical* as in [16, 21] where there is an implementation of operational parts with tools.

Abstraction can thus be classified as follows:
*Abstraction*: ENUM: {conceptual, technical}

### 5.3 Formalism facet

Generally speaking, representation formalism is a set of syntactic and semantic natural language, semi-formal such as diagram [17] or completely formal [15, 16].

A formal formalism is required to support the verification of the expected properties of the process model and validation of the process model using, for instance, simulation or enactment techniques. The use of informal notations has made it difficult for process models to be followed systematically. Formal or semi-formal formalism make these efforts considerably more effective as a formal formalism is necessary for providing automatic enactment support.

In the context of SME, the presented components have to be retrieved, assembled, tailored, and customized later and, hence, it is important that the representation approach will be formal or at least semi-formal.

The formalism facet helps classifying SME by one of the three values of the following enumeration:
*Formalism:* ENUM {formal, semi-formal, informal}

### 6. The Development view

The development view deals with two specific issues: the process of constructing component method, and the enactment of process as the SME methods are carried out with an aim of assisting the application engineers. The environment to offer assistance to the process in its execution course thus forms part of the problems whose solutions are provided by the development view. Three facets allow covering these aspects: *Flexibility, knowledge construction, knowledge organisation.*

### 6.1 Flexibility facet

Traditional methods (also named *rigid* methods) follow a static approach, which consist in prescribing entirely and statically the method, whereas SME methods use a contingency approach which consist in defining contingency factors defined on an application development. This is strongly related to the library of method components which must be enriched by the specific projects experiments.

[26] proposed a spectrum to organize the engineering methods approaches according to their degree of flexibility towards a new situation. The methods are organized on a scale of flexibility varying of "low" to "high". At the "low flexibility" level are the rigid methods while, at the "high flexibility" level, we find the SME methods. They are represented with the two last types of this spectrum:

- either the method engineer performs operations that have to be carried out on the original methodologies in order to create a new one, process that we will call *Customization*, as in [21] or

- he refers to methodology components, including their retrieval and assembly, process called *Modularity,* as in [3, 15]. In this last case, each component is usually treated as a closed unit that cannot be modified, while transformation and gluing parts between the components can be added in order to create "consecutive" methodologies.

This typology may be captured by a facet called *Flexibility* witch classifies the methods in two distinct categories.

*Flexibility*: ENUM {customization, modularity}

### 6.2 Knowledge construction techniques facet

The traditional knowledge construction is the expression of the application engineer experience. As long as this experience is not formalized and that a basic available knowledge does not constitute an available part for the various applications, one can say that this knowledge is the result of a *ad-hoc* construction technique [21]. This has two major consequences: ignorance in the way in how was carried out the construction and dependence on the field of expertise. If this knowledge must be independent of the expertise field and rapid to built, it is then necessary for construction techniques based on the experiment to use more *formalized* techniques [20].

The techniques of knowledge construction can thus be classified as follows:
*Knowledge Construction*: ENUM {formalised, ad hoc}

### 6.3 Knowledge organization techniques facet

The knowledge used during SME construction can be stored in library or *repository* to be reused later. Those provide the basic functions for the management of a components repository. As these libraries can contain a large number of components, they generally offer research techniques, as indexation techniques or the use of keywords.

Other approaches, in addition to the component extraction formalism, have an *organisational process* which helps to manage the knowledge coherence. The organization processes thus allow managing this problem in a more formal way [15, 20].

The knowledge organization technique can be classified as follows:

*Knowledge organisation*: SET (ENUM {repository, organization process})

Figure 2 summarizes the views and facets of the framework presented:

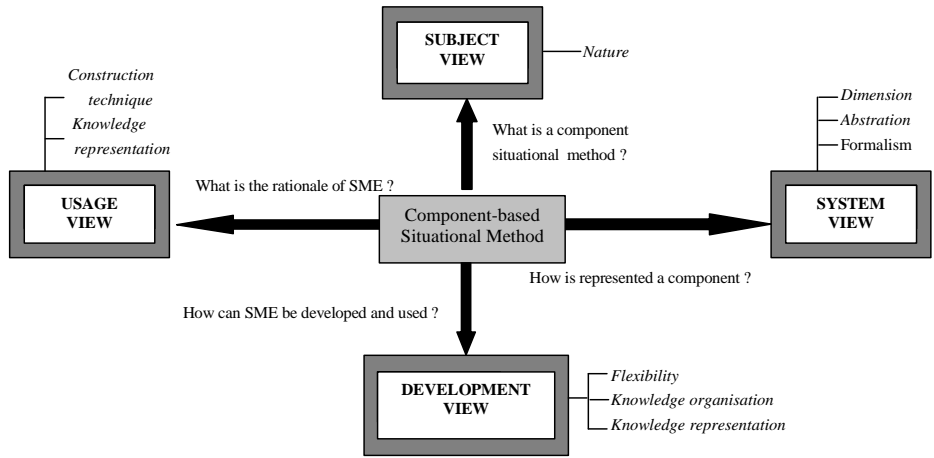

**Fig. 2:** Views and Facets of SME framework

This framework is used in the following section to evaluate a panel of SME approaches.

## 7. Review of SME approaches according to the framework

We propose a review of eight component-based SME approaches. We choose our method panel in the set of the most recent approaches and with the intention to offer a more complete study of the different views and their facets.
The aim is, firstly, to get a 'big picture' of the SME research area and to help understanding the achievements gained from currently developed SME based approaches in the literature. It is, secondly, to check the framework against eight SME approaches.

### 7.1 SME approaches

#### 7.1.1  Method configuration approach

[20] proposed a meta-method called Method for Method Configuration, which is based on the concept of Configuration Packages and Configuration Templates. These concepts are used to configure methods following the specificities of a project while creating reusable assets. Method configuration uses a specific base method as a basis for creating specific configurations. The reusability advantage is obvious since pre-made configurations can be used over and over again. Hence, there is no need to perform a complete method assembly or method configuration for each new project. Experiences

can be gathered and reused more efficiently since they can be attributed to coherent set of prescribed actions common in the organization, rather than to context-free actions.

### 7.1.2  Process Configuration approach

[21] proposes an approach called process configuration that tells how to create a project-specific methodology from an existing one, taking into account the project circumstances. The idea that lies behind is: for each individual project a specific process configuration (project-specific methodology) is create. This is done by selecting component from methodology that has been specifically designed for the organization and thus reflects the actual ways of working in the organization (base methodology). The configuration is done by processing the rules (in the engine) that are part of the base methodology. The rules define, for each methodology component, in what circumstances (project situations) its use is compulsory, advisable or discouraged.

### 7.1.3  Method extension approach

This approach [18] guides the method engineer by providing extension patterns that help identifying typical extension situations and provide advises to perform the required extension. In the extension-based, there is two ways to extend a method: directly through the pattern-matching strategy or by using some generic knowledge related to the domain for which the extension is to be done through the path select a meta-pattern, extend a method with the pattern-based strategy. The former help to match extension pattern stored in a library to the extension requirements whereas the latter select first, a meta-pattern corresponding to the extension domain and then, guides the method extension by applying the patterns suggested by the meta-pattern. Both way-of-working use a library of extension patterns but do it in different ways.

### 7.1.4  Method chunks approach

This approach [3] for assembly-based SME aims at constructing a method 'on the fly' in order to match as well as possible the situation of the project at hand. It consists in the selection of method components (called method chunks) from existing methods that satisfy some situational requirements and their assembly. This approach is requirements-driven, meaning that the method engineer must start by eliciting requirements for the method. Next, the method chunks matching these requirements can be retrieved from the method base. And finally, the selected chunks are assembled in order to compose a new method or to enhance an existing one. As a consequence, the three key intentions in the assembly-based method engineering process are: specify method requirements, select method chunks and assemble method chunks.

### 7.1.5  Application-based Domain Modeling (ADOM) approach

ADOM [17] is a domain engineering approach and uses the standard notation of UML 2.0 [27]. This approach is a visual methodology for managing representing, retrieving, customizing and tailoring situational method components. ADOM allows to express different types of methodologies and their components, their associated characteristics and values, their pre and post-condition and other component-related requirements, such as mandatory participants, recommended participants, triggers, etc. The structure and guidelines of components are described within the domain layer of ADOM, while their instantiations, which specify particular situational methodologies, are defined in the application layer.

### 7.1.6   Evolution-Driven (or Paradigm-Based) approach

This approach [15] uses meta-modelling as an underlying method engineering technique. The hypothesis of this approach is that the new method is obtained either by abstracting from an existing model or by instantiating a meta-model. Meta-modelling is known as a technique to capture knowledge about methods. It is a basis for understanding, comparing, evaluating an engineering method. A new methodology is then created by first constructing a product model and then a process model. There is different strategies available to construct both product and process model.

### 7.1.7   OPEN Process Framework (OPF) approach

The OPEN Process Framework [12] uses a meta-model to generate method components that are stored in a repository. OPEN offers a set of construction guidelines that are considered to be part of existing methodologies used to construct new methods. The OPF meta-model is composed of five main meta-classes [29] [30]: Stage, Producers, Work Units, Work Products and Languages; a method component is produced as soon as a meta-class is instantiated. An OPEN guideline helps method engineers both to instantiate the meta-model element to create method components and to select the best method components (from the repository) in order to create the situational method.

### 7.1.8   FIPA (Foundation for Intelligent Physical Agent) approach

FIPA [31] entered the IEEE computer Society Standards Committee with the mission of promoting agent-based technologies and the interoperability of agents with other technologies. FIPA defines the method fragment with a process meta-model [32, 33]. In this model, a process is composed of a set of activities performed by some active entities whose task is to produce a well-defined state of an Artefact as input/output. A process is defined as strongly oriented to the production of products. As a result, a method component [34] is defined as a reuse part of a design process composed of two elements: the structure of the product and the necessary procedures to construct this product [28].

## 7.2   Review of component-based SME approaches according to the framework

In this section, we propose a review of eight component-based SME approaches. Table 1 evaluates these eight approaches with respect to the four view defined facets and the four following sub-sections give more details about it, each one corresponding to a specific view of the framework.

**Subject View**
The subject view contains only one facet concerning the SME objectives. We can notice that almost all the approaches aims at documenting methods through well-defined fragments. Their strength resides in the effort of specification with regards to the elements a method is made of (tasks, activities, resources, etc.). In the method chunks approach, the focus is on the operators provided to allow a new combination of existing process fragments and on mechanisms to evaluate the similarity among them. Finally, five over the eight studied approaches focuses on method fragmentation for project team members, to provide them with guidelines which are to be reused while performing their daily task. We see here that only the Method chunks approach is addressing the particular aspect of the component retrieving with a formalized evaluation strategy. However, we think that the retrieval and selection of a component is a very important issue of the SME field and that a particular attention had to be drawn on it.

**Usage View**
The objective for the methods engineers, in situational approaches, is to make methods completely flexible and situation adaptable. This is possible with the components that enrich method library or repository and are reused for method construction. Thus, method chunks use directives and signatures. Its method construction technique is done by *instantiation* and chunks *assembly*. The following methods use the same technique: Evolution-Driven, ADOM-UML and OPEN. Method configuration and processes configuration build their method through the technique of *reduction* and *extension* and they propose a combination of the cancellation and extension operators. All of this show that the construction techniques are often combined, which increase the flexibility of the SME approaches.

The components representation of these methods varies. Thus the method extension uses extension *patterns* and the FIPA method defines its components like a set of activities. As knowledge representation model, method configuration uses packages and templates as *method components* and process configuration uses *process components* which are then to enrich by a set of rules which define how the component has to be used. More than half the approaches studied use a chunk knowledge representation, which allows to describe more effectively the component.

**System View**
Six of the studied methods (chunks, extension, OPF, Evolution-Driven, FIPA, ADOM) integrate two aspects of the method fragment, the *product* and the *process*, so they represents a portion of process together with its related product(s). Process configuration tends to bring the construction *process* closer to its users by providing facilities for managing the rules. Method configuration is based rather on the *product*.
Regarding the abstraction, we notice that some methods (Process Configuration, OPF, Evolution driven, ADOM) define their fragments as *technical* fragment i.e. in the form of tools. ADOM-UML has to develop a supporting CASE tool for managing the activities. On the same way, Evolution-Driven develop the LyeeALL CASE tool in order to generate programs, as a set of well-formatted software requirements are given. OPF use the tool

OPENPC (OPEN Process Construction) that use the OPEN repository of methodological components (firstly conceived for the development of directed objects but used widely for other applications).

Three methods (configuration, chunks and evolution-Driven), have a *formal* representation approach. These approaches deals with the definition, the representation, the cataloguing of components according to different features, the retrieval of the most appropriate ones, and the customization and tailoring of them to complete methodologies that best fit a given situation. In these approaches, component representation and cataloguing are very important activities. The others approaches are *semi-formal*.

**Development View**
Regarding flexibility, four approaches (method extension, method configuration, process configuration) enable all their components to be specialized, adapted and *customized*. These operations create new components that can be modified as requested by allowing specification of gluing and transformation components, customization parts. They start with a particular basic method as initial point of departure, then configure them with different reusable components. In that case, there is no assembly but rather a configuration from different parameters or reusable components. In Method configuration, the configuration of a methodology is supported by configuration packages and configuration templates which present reusable assets that can be used in particular software development situation. In process configuration, each process component or components is supplemented by a set of rules that define when to use the component. Method extension uses the patterns as reusable components to configure the method. On the other hand, the other methods (ADOM-UML, OPEN, method chunk, Evolution-Driven, FIPA) use *modularity* construction strategy's which focus on consistent and congruent method modules. Project-specific methodology is created from fragments that might come from different methodologies. These approaches design their final approaches starting from a set of different and reusable modules to assemble them. This illustrates that authors do not favour one approach to the other, they either use customization or modularity.

All methods use a *library* to organize the components. Some of them also use an *organization process* to manage the coherence. Thus, method extension and Evolution-Driven use the process organization based on the "Map" process of [35] to organize their components. Method configuration proposes an organization based on three *repositories* of components (characteristic, configuration packages and templates). This is showing that the use of a library is required when using an SME approach, as all the components have to be stored somewhere. However, the use of an organization process is not always offered. This may be an issue that authors should work on.

| Views | Subject | Usage | | System | | | Development | | |
|---|---|---|---|---|---|---|---|---|---|
| Facet | Nature | Construction technique | Knowledge representation | Dimension | Abstraction | Formalism | Flexibility | Knowledge Construction | Knowledge organization |
| Process Configuration | Documenting + Reuse Guiding | Extension + Reduction | Chunk | Process | Technical | Semi-formal | Customization | Ad hoc | Repository |
| Method configuration | Reuse Guiding | Extension + Reduction | Fragment | Product | Conceptual | Formal | Customization | Formalised | Repository + Organization process. |
| Method Extension | Documenting + Reuse Guiding | Instantiation + Extension | Pattern | Product + Process | Conceptual | Semi-formal | Customization | Formalised | Repository + Organization process |
| OPEN Process Framework | Documenting, + Reuse Guiding | Instantiation + Assembly | Chunk | Product + Process | Technical | Formal | Modularity | Formalised | Repository |
| Method chunks | Documenting + Reuse Guiding + Retrieving and evaluating | Instantiation + Assembly | Chunk | Product + Process | Conceptual | Formal | Modularity | Formalised | Repository |
| FIPA | Documenting + Reuse Guiding | Assembly | Chunk | Product + Process | Conceptual | Semi-formal | Modularity | Ad hoc | Repository |
| Evolution-Driven | Documenting + Reuse Guiding | Instantiation + Assembly | Chunk | Product + Process | Technical | Formal | Modularity | Formalised | Repository + Organization process |
| Method ADOM | Documenting | Instantiation + Assembly | Fragment | Product + Process | Technical | Semi-formal | Customization | Formalised | Repository |

**Table 1**⊥ Review of SME methods

# 8. Conclusion

Our study has shown that component-based SME approaches are very complex, multi-dimensional entities. They cannot be treated adequately with simple predicate based classification techniques. Rather, the need is for a 4-dimensional framework for a component-based approach to be well described.

Every view is itself multi-faceted. Some facets have been proposed by other researchers earlier, others have been introduced by us here. We believe that we have incorporated in our proposals a comprehensive set of facets which cover all the dimensions of our framework.

Through the notion of a view and a facet, we are able to successfully capture the global view and the more detailed view of a component-based SME approach. In this way, the individual characteristics of these approaches are captured within the larger view of SME nature, component management policy, component use and knowledge representation and construction.

Applying the framework on eight recent approaches shows that they all share some of the properties that characterise component-based SME methods. However, they differ in a lot of the selected parameters and their application to this framework allow a precise inventory of their differences.

One of our objectives for our further researches is to review more of the existing SME approaches in order to apply our 4-dimensional framework on a panel as complete as possible. This will allow us to test the validity of our framework and maybe to identify more facets to compare more effectively the methods. Moreover, discussion with other SME approaches authors will help to check the validity, or the invalidity, of our facets.

The main perspective of this work is to identify the real key facets of SME in order to identify reusable components *from* these construction approaches. As a result, components would be of two types, either a capture of method knowledge or a capture of method construction knowledge. This will offer the possibility to the method engineer to reuse them in order to create a new approach to construct SME methods, perfectly adapted to his way of working.